\documentclass[12pt,preprint]{aastex}

\begin{document}

\title{The evolution of the galaxy luminosity function in the rest frame
blue band up to $z=3.5^1$}

\author{F. Poli$^2$, E. Giallongo$^2$, A. Fontana$^2$, N. Menci$^2$, G. Zamorani$^3$, M. Nonino$^4$, 
P. Saracco$^5$, E. Vanzella$^6$, I. Donnarumma$^2$,
S. Salimbeni$^2$, A. Cimatti$^7$, S. Cristiani$^3$, E. Daddi$^6$, S. D'Odorico$^6$,
M. Mignoli$^3$, L. Pozzetti$^3$, A. Renzini$^6$}

\begin{abstract}

We  present an  estimate of  the cosmological  evolution of  the field
galaxy luminosity function (LF) in the rest frame $4400$ {\AA}  B-band
up to   redshift $z=3.5$. To this  purpose, we use  a composite sample
of $1541$  $I$--selected galaxies  selected down to $I_{AB}=27.2$  and
$138$ galaxies selected  down to   $K_{AB}=25$  from  ground-based and   HST
multicolor  surveys, most   notably the  new  deep    $JHK$ images  in
the Hubble Deep  Field  South (HDF-S)    taken
with  the   ISAAC   instrument  at   the   ESO-VLT  telescope.  About
$21\%$ of  the  sample     has   spectroscopic     redshifts,      and
the  remaining    fraction well    calibrated  photometric  redshifts.
The resulting   blue  LF     shows  little density evolution  at   the
faint end    with respect   to  the    local  values,   while  at  the
bright  end   ($M_B(AB)<-20$)     a  brightening    increasing    with
redshift      is  apparent   with  respect    to   the     local  LF.
Hierarchical CDM     models overpredict   the    number     of  faint
galaxies by    a   factor  $\sim    3$   at $z\simeq    1$.   At   the
bright   end   the     predicted    LFs    are         in   reasonable
agreement   only at    low    and intermediate    redshifts  ($z\simeq
1$),      but fail       to reproduce    the    pronounced brightening
observed    in  the    high   redshift    ($z\sim   2-3$)   LF.   This
brightening      could mark    the  epoch   where  a   major      star
formation activity  is present      in the galaxy   evolution.

\end{abstract}


\altaffiltext{1}{Mainly based on service mode observations collected at the European
Southern Observatory, Programme 164.O-0612}

\altaffiltext{2}{INAF, Osservatorio Astronomico di Roma, via Frascati 33,
I-00040, Monteporzio, Italy} 

\altaffiltext{3}{INAF, Osservatorio Astronomico di Bologna, via Ranzani 1,
I-40127, Bologna, Italy} 

\altaffiltext{4}{INAF, Osservatorio Astronomico di Trieste, via G.B. Tiepolo 11,
I-34131, Trieste, Italy}

\altaffiltext{5}{INAF, Osservatorio Astronomico di Brera, via Brera, 28,
20121, Milano, Italy} 

\altaffiltext{6}{European Southern Observatory, Karl-Schwarzschild-Strasse 2,
D-85748, Garching, Germany} 

\altaffiltext{7}{INAF, Osservatorio Astronomico di Arcetri, Largo E. Fermi 5,
I-50125, Firenze, Italy}

\section{Introduction}

The  evolution  of  the  luminosity distribution  of  field   galaxies
and  its correlation with colors constitutes the baseline for studying
the link  between  the stellar  processes  occurring  in galaxies  and
the dynamical  properties   of  their host  dark matter    (DM) haloes
which are   thought to  be described   by the  hierarchical clustering
theories. To this aim,  observations should i) sample a wide range  of
rest-frame wavelengths from the UV down  to the optical and NIR  bands
to derive  the contribution  of the  different stellar  populations to
the total mass and star  formation rate; ii) be extended    to   faint
magnitudes,    since  feedback effects --  regulating     the   amount
of    gas    which  can    actually cool    and form   stars --    are
effective       in  depleting       the  gas    content    of galaxies
in      the        shallow     potential       wells,     and    hence
appreciably affect   the faint  end of  the   LF;  iii) be  performed
in   the  same  rest-frame    band    to   avoid   the    mixing  (the
degeneracy) between  redshift   and   wavelength  dependent   effects,
possibly  reducing   the effect   of dust  extinction. Fulfilling  the
above requirements up to high  redshifts with a homogeneous sample  is
not   an   easy  task.    At  present,  the   available   results  are
scattered  in different bands,  depending on  $z$, and  have different
extensions at  faint luminosities.

At $z\approx 0$, recent results in the B band from  
redshift surveys  (Zucca  et al.  1997,  Folkes  et al. 1999,  Blanton
et  al.   2001)  have  shown that  the   local  population   of  field
galaxies has   a LF  with  a  fainter characteristic  magnitude and  a
faint slope steeper than previously  found. At intermediate  redshifts
($z\simeq0.75$), spectroscopic surveys  (CFRS Lilly et al  1995;
Autofib, Ellis  et al. 1996;  CNOC2, Lin  et al.  1999)
found a  steepening  of the faint end  LF
with increasing $z$  in  the total LF, mainly   due  to
the   contribution by  later   type galaxies.  To
investigate  the  faint   end  shape  and  evolution   of   the  LF at
higher $z$,  the statistics   must be extended well beyond  the  limit
of  spectroscopic    surveys.  The   photometric redshift   technique,
(Connolly  et al.  1997, Giallongo et  al.   1998, Fernandez-Soto,
Lanzetta,   \&  Yahil  1999,  Fontana   et   al.   2000), allowed   to
otain  a  first estimate  of   the  rest-frame  B LF    at  $0<z<1.25$
(Poli  et al.   2001) using  a composite  deep   multicolor sample  in
the   Hubble   Deep    Fields  (HDF)   North   and   South  down    to
$I_{AB}=27.2$. The lack of deep NIR images prevented us to follow  the
evolutionary behaviour of  the LF  in  the same  rest frame B band  to
$z>2$. 

In this paper,  we  present the  results from deep  ISAAC/VLT IR  data
in   the  HDF-S,  which   are   providing  a     complete   sample  of
galaxies   down      to  $K_{AB}=25$, thus   giving     us  the  first
chance to  assess   the  evolution of  the  galaxy LF   in  the   same
rest   frame $4400$   {\AA}    B-band from redshift $z=0$  to $z=3.5$.
Throughout   this    paper  we    adopt   an   $\Omega_{\Lambda}=0.7$,
$\Omega_{M}=0.3$ and $H_{0}=70$ km $s^{-1}$ Mpc$^{-1}$ cosmology.

\section{The data sample}

The  dataset  we  have  analyzed consists  of  a  composite  sample of
galaxies covering, with  similar  depth,  a range  spanning  from  the
UV  down   to  the  optical/NIR  wavelengths  (see  Tab.  1).  A first
photometric   catalog  has  been  extracted  from  the  HDF-S  dataset
where   HST-WFPC2   optical   images  (Casertano  et  al.  2000)  have
been combined with  deep ESO/VLT NIR  images  in  a  $4.22$ $arcmin^2$
sky area  (these data are in  common with the FIRES   survey,  Labb\'e
et al. 2003).  The infrared images   reach formal  5  sigma limits  in
an   aperture   of  $1.2''$  at    $J_{AB}  =25.3$,  $H_{AB}=24$,  and
$K_{AB}=25$.    The    procedures   adopted  to   extract  the   final
multicolor   catalog   have  been   optimized   to     obtain accurate
color  photometry   for  reliable photometric    redshifts, as
described in  Vanzella  et  al (2001).  The   full
catalog    will be     presented  in   a  future    paper. From   this
catalog  we    have selected   245  $I_{AB}\le    27.2$    sources  to
build  the  LF  in the  $0.4<z<1$ range,   and  113    $K_{AB}\le 25$
sources    to  sample   the  LF    in     the    same      rest  frame
wavelength    but    at  higher  redshifts, $1.3<z<3.5$.   To  improve
the  statistics    at  the  faint end    of the   LF  we   added   259
$I_{AB}\le  27.2$ at $z\leq  1$ and 18  $K_{AB}\leq     23.5$ galaxies
at     $1.3<z<3.5$  from  the  HDF-N   catalog  of   Fontana  et   al.
(2000).  To   improve the  statistics also  at the      bright end  of
the    LF,  we  added  a  sample   of   galaxies  selected     in  two
fields  centered     around  the     QSO  0055-269   and   in     the
Chandra Deep   Field  South (Giacconi   et   al.  2001)   for  a  total   of
$\simeq 68$  $arcmin^2$,  that     were  used      to   select     the
targets     for the   so called  K20 ($K_{AB}\le 21.9$)  spectroscopic
survey (Cimatti et  al. 2002).

Simulations show   that    the    K20 sample
and  HDFs    are   complete   (to   more       than  95\%)    down  to
$I_{AB}=24.5,25$ and $I_{AB}=27.2$, respectively. Following  the  procedure
presented in     Poli et   al. (1999),   we  separately  estimate {\em
total}  and {\em  color} magnitudes  with two  different recipes.   To
obtain   the   total   fluxes   of   the   objects   in   the   chosen
reference   band,  SExtractor  Kron   magnitudes  (Bertin   \& Arnouts
1996) have   been adopted    for bright    sources. For  faint sources
we    estimated   aperture  magnitudes      with    aperture
corrections  computed  estimating  the   flux   losses   outside   the
adopted aperture with respect   to  the Kron magnitudes for relatively
bright sources.  The reliability   of  the    correction adopted   has
been  tested  by   means   of simulations    as  performed    in  Poli
et   al.      (2001).      No  appreciable   systematic    losses   of
the total    flux    for   the   faintest  sources    were  found    a
part     from  the   well   known   losses  by    5-10\%   typical  of
the     Kron  magnitudes.   To    cope    with    the    inhomogeneous
quality  of   the  different   images, we  produced a  set
of seeing--corrected {\em  color}   magnitudes that  have  been   then
scaled  to  total    magnitudes in   all   the  bands
(Vanzella et al. 2001).  This    ensures that  colors  -  used   to
estimate  the    redshifts -    are measured in   the    same physical
region  at any wavelength,   independently of the seeing FWHM  present
in each image.

The     final     sample   includes    1541     $I$--selected      and
$138$     $K$ -selected  galaxies   (see   Tab.   1).    Of     these,
$337$     and     $23$,  respectively,      have    spectroscopic $z$,
taken      from  existing spectroscopic    surveys  (Cohen et  al  for
the  HDF-N; Vanzella   et   al.    2002 for   the HDF-S;   Cimatti et
al. 2002   for  the  K20 sample).    For  all   the  other   galaxies,
reliable   photometric      redshifts  have      been derived  with  a
homogeneous     technique,  described  in   Fontana  et  al 2000  (for
the  HDF-S  sample)  and   in   Cimatti et   al 2002  (for     the K20
survey).  The  model  adopts  several  galaxy  ages  and exponentially
declining  star  formation histories, and  includes
Lyman   absorption     from  the   intergalactic   medium   and   dust
absorption with different  amounts  up to   $E(B-V)=1$,    using  both
SMC and   Calzetti  extinction    curves. In    all     the   samples,
the      relative   accuracy  is   $(z_{spe} -z_{phot})/(1+z_{spe})
\lesssim 0.05$.

\section{Bulding the observed LF}

Our combined  sample  allows  to probe   the behavior  of  the  $4400$
{\AA}  rest frame  LF  in  a   continuous way   from  $z\sim  0.4$  up
to $z\sim    3.5$.  For  galaxies  at   $z<1$ we   used  our  $I-$band
selected  sample.  Indeed,  in    the redshift  interval $0.7<z<1$ the
4400 {\AA}  rest frame  wavelength is  within  the  I  band,   and the
LF  includes    all the   galaxies with     $m[4400(1+z)]=  I_{AB}\leq
I_{AB}(lim)$ where $I_{AB}(lim)$ is 24.5, 25 and 27.2 for the  shallow
and  deep  samples,   respectively.    In   the      lower    redshift
interval   $0.4<z<0.7$     the wavelength  corresponding to  4400(1+z)
lies   shortward of  the I    band but  still we    adopt the    same
threshold in  $m[4400(1+z)]$. This   implies that  some  galaxies from
the original  $I_{AB}$   limited samples are   excluded from the    LF
since     they     have   a     red     spectrum   and    consequently
$m[4400(1+z)]$ are fainter than our adopted threshold.   We  have   in
this way extracted  a  complete  subsample   of   galaxies  having   a
continuum  magnitude  threshold at   4400 {\AA}  in    each   redshift
interval,  independently     of  their  color     (assuming   galaxies
in   this    redshift   interval   to    have   $(R -I)_{AB}\geq  0$).
The  same procedure  has  been adopted   at higher  $z$  using  the $K
-$band   selected  sample. In   the $1.3<z<3.5$   interval,  the  4400
{\AA}  rest frame   wavelength   is  within  or shortward of  the    K
band,    and    the    LF    includes    all    the    galaxies   with
$m[4400(1+z)]=K_{AB}\leq  K_{AB}(lim)$.  Finally,    the  rest   frame
absolute   AB magnitude   $M_B(AB)$  has   been derived    from   the
theoretical  best  fit  spectral     energy  distribution   for   each
galaxy. The  same   fit   provides the  photometric redshift  for each
galaxy.  In   this    way the   K-correction  has   been  derived  for
each   galaxy from the  interpolation between the observed  magnitudes
using  the  best  fit spectrum.  Uncertainties  in  this interpolation
produce on average  small ($\lesssim 10\%$)  errors in the  rest-frame
luminosities  (Ellis  1997;  Pozzetti et  al.  2003).  The rest  frame
luminosity is derived without any correction for dust absorption.

To   estimate   the   observed   LF   we    have   applied   to    our
composite  sample an extended  version of  the    $1/V_{max}$
algorithm.  Our  combination  of  data  from  separate   fields,  with
different  magnitude  limits, has   been    treated  computing,    for
every     single     object,     a    set     of   effective  volumes,
$V_{max}(j)$, for each  $j$-th   field  under analysis. For a    given
redshift interval  $(z_1,z_2)$, these   volumes are enclosed   between
$z_1$  and $z_{up}(j)$,  the  latter being   defined  as  the  minimum
between $z_2$   and  the  maximum  redshift   at  which  the    object
could have  been observed  within  the magnitude  limit  of  the   $j$
-th  field.  

The  Poisson error  in  each LF  magnitude  bin was  computed adopting
the   recipe  by  Gehrels  (1986)  valid  also  for  small numbers. To
compute   the  uncertainties  in  the LF due  to    the    photometric
$z$  estimates,   we    assumed    the    typical     r.m.s.     found
comparing   photometric     and        spectroscopic   $z$      (where
available) in  the   HDF-N  and K20 samples.   The adopted      values
range  from  $\sigma    _z    \sim      0.08$   at   $z\sim   1$    to
$\sigma  _z \sim    0.16$     at    $z\sim     3$    (Fontana   et al.
2000,   Cimatti      et   al.   2002).  We have  generated  from   the
original   $z$   photometric  catalogs  of I-selected   or  K-selected
galaxies,     a      set   of   50   catalogs  of  random  photometric
$z$  using the above   r.m.s. uncertainties to     compute       a set
of 50  LFs. The  derived      fluctuations in each      magnitude  bin
resulted smaller   than  Poissonian  errors and    have    been  added
in  quadrature.  Due to  the  limited area  of  each survey,   density
fluctuations among the fields could be larger than Poissonian  errors.
For this  reason both  such uncertainties  have been  computed in each
magnitude  bin  of  the LF   at $z<1$.   At $z>1$   the LF   has been
derived  from  the deep  K  selected  HDF-S   sample   and thus   only
Poissonian  errors have  been computed.

A   complementary  way  to  study   the   evolution   of  the   galaxy
population is    to apply  a maximum   likelihood   analysis  assuming
a    Schechter     parametric  form  $\phi$   for  the   LF  (Sandage,
Tammann  \&   Yahil  1979). 
In the case of  our composite survey, that  is based on $j$  magnitude-limited
fields, the maximum likelihood $\Lambda$ writes (Marshall et al. 1983)

\begin{equation}
\Lambda = \prod_{i=1}^{N}\frac
{\phi(M_i)dzdM}{\sum_j \omega(j)\int_{z_1}^{z_2}\frac{dV}{dz}dz
\int_{-\infty}^ { M_{lim}^{j}(z) }\phi(M)dM}
\end{equation}

$\omega(j)$ beeing the field  area in steradians and  $N$ the total number  of
objects considered in the  composite sample. $M_{lim}^{j}(z)$ is  the absolute
magnitude   value  corresponding,   at  the  given  redshift  $z$,   to    the
magnitude limit  of  the  $j$-th  survey.  The   value   of    $\phi^{*}$   is
then   obtained  simply     imposing      on    the   best     fit     LF    a
normalization such that the total  number of  galaxies of the combined sample
is reproduced.

\section{The evolution of the LF in the rest frame 4400 {\AA} band}

The  resulting  LFs  in  different redshift   intervals are  shown  in
figure  1,  where  in    each   panel  the  local    (i.e.  $z\lesssim
0.1$)  observed  blue LFs  from  the  Sloan (Blanton et  al.     2001)
and 2DF (Norberg et al. 2002) surveys have been shown for  comparison.
First   of all   we note   that the   LFs derived   from the   present
$I$-selected composite  sample  up  to   $z=1$   show  little  density
evolution   at the  faint end   respect    to the  local values    and
the   density  remains  approximately  constant   for  $M_B(AB)\sim   
-16$. A small underdensity is apparent at $0.7<z<1$ and  $M_B(AB)\sim 
-19$ but this appears of the same order of the differences between the
two  local  LFs.  At  the bright  end  a  brightening  increasing with
redshift  is  apparent with respect    to  the local  LF.  Indeed,  as
shown       in  Table  2,    assuming          a    Schechter    shape
for  the      LF,   the      characteristic      luminosity    $M_B^*$
increases     from   $  -21.3$    to   $   -21.9$    in    the   range
$z=0.4-1$.

It  is    interesting to     compare  our    LF at      $z\lesssim  1$
with that derived from     the COMBO-17    survey  (Wolf      et   al.
2003)  based   on photometric redshifts  derived   from  a
shallower   sample ($R<24$),  but  from  a  larger    area      of   1
deg$^2$.      They     derive           for   the   overall     sample
characteristic    luminosities           $M_B^*(AB)=-21, -21.9, -22.2$
(for   $H_0=70$ km  $s^{
-1}$   Mpc$^{-1}$)    and    slopes    $\alpha   =-1.1,-1.6,-2$     at
redshifts $z=0.5,0.7,0.9$ respectively. Formally the slopes derived by
the  COMBO-17  survey  are  not  statistically  consistent  with those
obtained by our samples at similar $z$. However, given the
significantly different limiting magnitudes of the two samples, it  is
difficult to  compare  with each  other the  two sets  of
slopes.  In any  case, both  samples suggest  a steepening  of the  LF
coupled with  an   increase    of  $M_B^*$  from
$z\simeq 0$ up to $z\sim 1$. 

At   $z\sim    2-3$ our LF  still shows  a  brightening with   respect
to  the       local value  which appears    to  be    due to  enhanced
star  formation    activity    for   the    brightest galaxies.   This
can  be  derived    from     figure     2         where we        plot
the   correlation    existing    between    the  observed   rest frame
blue    and   the     de-reddened    UV    1400    {\AA}  luminosities
derived    from   the spectral  models fitting the NIR/optical  colors
of  each  $K$  selected     galaxy.   The  median  reddening   for the
galaxies in  Fig. 2  is  $E(B  -V)\sim   0.1,0.2$    for   an  SMC  or
Calzetti    extinction  curves, respectively,    in  agreement    with
Shapley  et al.   (2001). We  have    also checked   the    robustness
of such  a result   using the correlation (Meurer et al. 1999) between
the  rest-frame UV  shape (derived  from the  observed optical colors)
and the  UV   extinction,  a  procedure    less   dependent    on  the
estimated    age  of   the stellar population.

Thus,  the   correlation shown   in  Fig.  2   implies   that brighter
blue galaxies   have   on average   greater  star  formation  activity.
We note    in this    respect  that
67\%, 83\%,  67\%    of  the  brightest   galaxies with  $M_B(AB)<-22$
and  $z<1$,   $1.3<z<2.5$    and   $2.5<z<3.5$    have   spectroscopic
redshifts. Thus   the    amount       of  brightening  is     measured
with    good  accuracy  despite    the    small  number   of   sources
involved   at    the  bright    end.   We   also    note    that  some
galaxies  in  figure     2 contributing  to  the    LF are     clearly
red   with   a    small     UV  emission (see    also   Franx et   al.
2003). It  is  difficult to argue about   their nature  (old or   dust
reddened)   without  spectroscopic  information   (see  Franx  et  al.
2003). However these red galaxies  have in general  $M_B>-22$ and   so
they  have magnitudes  of the  order of $M^*$  or fainter and   do not
affect  the brightest end  of the  blue LF. They represent about  9\%
of the sample shown in  figure 2.

At the    faint   end   of the  LF, the  Schechter slope $\alpha$   is
not  well    constrained     at      these  redshifts.     The derived
slope       at    $z=1.3   -2.5$     is       $\alpha=    -1.8$    and
the     related         characteristic          luminosity          is
$M_B^*\simeq   -22.4$, but the   uncertainties    are  large.   Fixing
the    slope       to         the   lower       redshift         value
$\alpha=    -1.4$              we    derive    a        characteristic
luminosity          $M_B^*\sim  -21.9$.  Keeping  the  slope  fixed  a
further     brightening     by about  0.4 magnitude ($M_B^*\sim  -22.3
$)    is    apparent in  the redshift  interval      $z=2.5-3.5$  (see
Table 1). Similar results concerning the evolution of the blue LF  and
derived luminosity density have   been obtained in the    HDF-N NICMOS
field by Dickinson et al (2001, 2003). Our results are also consistent
with the rest-frame  V  band  LF  derived at $z\simeq   3$ by  Shapley
et al. (2001).

The above results are  compared with the semi-analytic  model by Menci
et   al.   (2002)  describing  the  galactic  merging   histories   in
the  framework  of hierarchical    models.    The  cooling,       star
formation      and  Supernovae  (SNe)    feedback    of  the  galactic
baryons  in  a   way   similar    to  other    semi-analytic    models
(Kauffmann  et  al.  1993; Somerville  \&   Primack  1999; Cole     et
al. 2000).  Stars  are allowed to  form  from the  cooled     gas with
rate $\dot     m_*    =  (m_c/    t_{dyn})({v/   200\,{\rm     km\,s^{
-1}}      })^{   -\alpha_*}$  where     $t_{dyn}$      is   the   disk
dynamical   time.      A  mass  $\Delta  m_h=\beta\,m_*$   is returned
from   the   cool to   the  hot  gas  phase    due  to    the   energy
feedback      by    type   II  SNe associated      to   $m_*$.
The   feedback         efficiency         is  taken      to         be
$\beta=  (v/v_h)^{\alpha_h}$.    The     values  adopted      for  the
parameters are    the    same    of   Cole     et al.   (2000)     and
allow  similar fits     to  the    local    B-band galaxy    LF    and
the   Tully-Fisher  relation, as  illustrated by Menci  et al. (2002).
However,  our   model also includes  a  physical prescription for  the
binary  aggregations  among satellite    galaxies in   common  hosting
haloes,    which    provides    a    somewhat      flatter  luminosity
function at   all $z$.   In  comparing   the observed   LF with   that
predicted   by  our   CDM  model  we  did    not correct  the observed
rest  frame  luminosities for dust   absorption but we   have included
the   reddening   effects    in   the   model.   The   dust absorption
has been  assumed  to  be proportional   to the  disk  surface density
and  metallicity as  detailed  in  Menci  et  al. (2002).  We  adopted
an   SMC  extinction   curve;  using    a  Milky    Way   or  Calzetti
extinction  curves  does not  affect  appreciably  the  resulting    B
band  luminosities.

The comparison shows    two main  features: {\bf  a)}   At  the  faint
end  of  the   LF the  excess   of  predicted  galaxies  persists   at
all   redshifts    (see      also Poli  et   al. 2001,  Cimatti     et
al.   2002b).    This  implies     that    some  feedback  could    be
effective   in     suppressing    the    star     formation  activity.
However,  since  further   increments  of    the  SNe  feedback  would
result  in    a   local  Tully-Fisher  relation    too  faint     when
compared     with  the    data, sources      of feedback   other  than
SNe    should     be    investigated. An     example   is  represented
by  the       feedback    provided  by  the    diffuse ionizing     UV
background        produced   by  QSOs. {\bf  b)}     At   the   bright
end  ($M_B(AB)<-20$),      the   theoretical  LFs     show      little
evolution      up    to    $z=3.5$, resulting  from      the   balance
between        the decrease   with   z of        the        number  of
massive   objects     and      the    increase   of  their        star
formation  and    hence       of their     blue  luminosity.
On  the     contrary, the    observed    LFs     show   an  increasing
brightening     with    redshift.   Even       if     at         $z<1$
the predicted     LFs       are        still  consistent   with   this
brightening,  at    $z>1$    they      fail     to      reproduce  the
observed   number of   bright galaxies. This discrepancy   would  then
enlighten     the      lack      in   hierarchical      models      of
sufficiently  high   star     formation     activity    at    $z=2-4$,
beyond the quiescent mode driven  by the progressive  refueling of the
cold  gas  reservoir  during  galaxy  merging   and  by  the   gradual
cooling  of the  available   hot   gas.   An  additional   burst  mode
triggered  by  galaxy interactions     in common     DM   halos  (e.g.
Somerville     \&  Primack  1999;    Poli    et   al.    2001)   could
contribute to solve   the problem. Indeed, the  rate of   interactions
peaks in    the above  redshift  interval  (Menci
et al. 2002).

\newpage

\centerline {Table 1.}
\centerline{The composite sample}
\vspace{0.3cm}
\centerline{
\begin{tabular}{llccc}
\hline
\hline
field  & maglim & arcmin$^2$ & Filters  & N \\
\hline
HDFS    & $I<27.2$ & 4.22 & UBVIJHK & 245\\
HDFS    & $K<25$ & 4.22 & UBVIJHK & 113\\
HDFN    & $I<27.2$ & 3.92 & UBVIJHK & 259\\
HDFN    & $K<23.5$ & 3.92 & UBVIJHK & 18\\
Chandra & $I<25$ & 35.6 & UBVRIZJK & 619\\
Chandra & $K<21.9$ & 24.6 & UBVRIZJK & 6\\
Q0055-269 & $I<24.5$ & 31.9 & UBVIJHK & 418\\
Q0055-269 & $K<21.9$ & 19.8 & UBGVRR$_W$IZJK & 1\\
\hline
\end{tabular}
}

\newpage

\centerline {Table 2. }
\centerline{Parameters of the Schechter function fits}
\vspace{0.3cm}
\centerline{
\begin{tabular}{lcccc}
\hline
\hline
z range   &$\alpha$&$M^*$&$\phi_*$&N \\
\hline
0.4-0.7    &$-1.33\pm 0.04$ &$-21.33\pm0.17$ &0.004 &787\\
0.7-1  &$-1.44\pm 0.04$ &$-21.88\pm0.19$ &0.002 &754\\
1.3-2.5 &$-1.80\pm 0.20$ &$-22.37\pm 0.39$ &0.0007 &96\\
1.3-2.5 &$-1.44$ &$-21.87\pm0.14$ &0.0017 &96\\
2.5-3.5 &$-1.44$ &$-22.30\pm0.31$ &0.0015 &42\\
\hline
\end{tabular}
}

\newpage

\begin{figure}

\epsscale{0.8}
\plotone{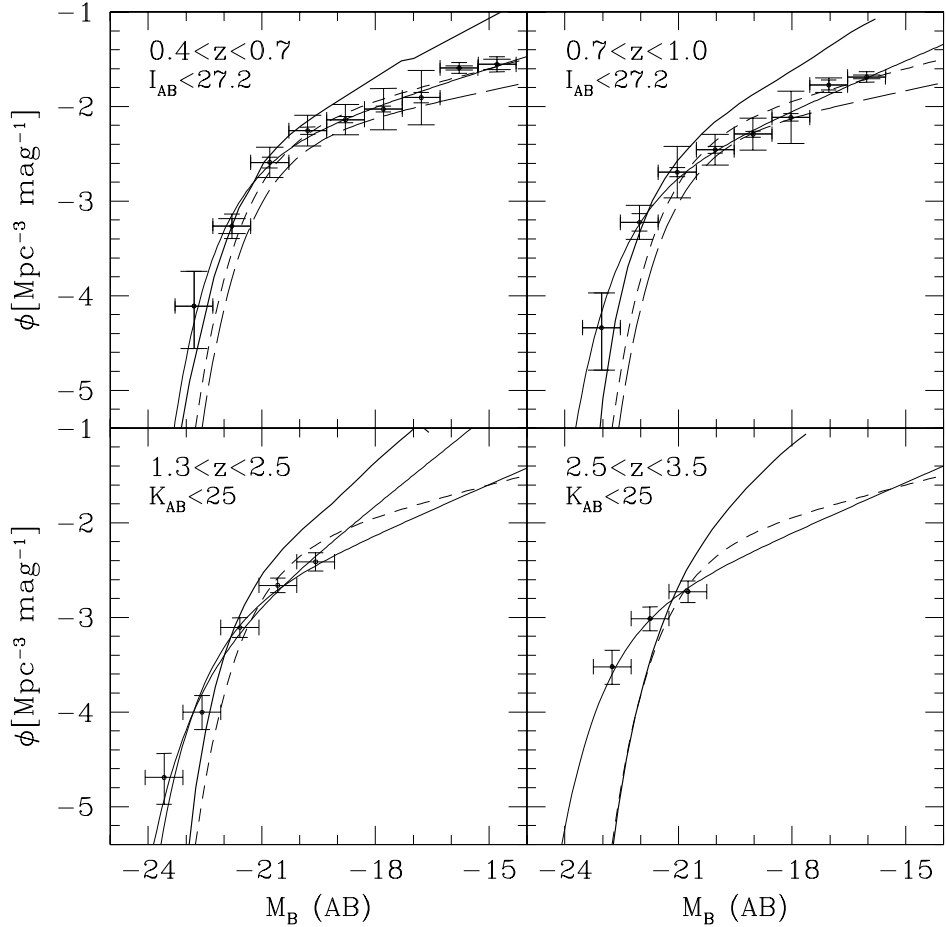}

\caption{ Blue   luminosity
function at  various  redshift  intervals.  Thick continuous curve  represents
the LF predicted  by the CDM  model of Menci  et  al. (2002).  Thin continuous
curve represents  the  Schechter  fit to the unbinned  data where the flatter slopes
in the lower panels result from the fit with $\alpha =1.44$. Short  dashed
curve is  the observed  local LF  from the  Sloan survey (Blanton et al. 2001)
where  $M^*=-20.65$  has  been   derived   from  the $g$  magnitudes  adopting
the following transformation  $M^*_B=g^*+0.32$ and scaling magnitude and
normalization to $H_{0}=70$ km $s^{-1}$ Mpc$^{-1}$. Long dashed curve is
the observed local LF from the 2dF survey. Larger vertical error bars
come from the r.m.s. field-to-field variance, smaller bars from the poissonian
statistics.
\label{fig1}}
\end{figure}

\newpage

\begin{figure}

\epsscale{0.8}
\plotone{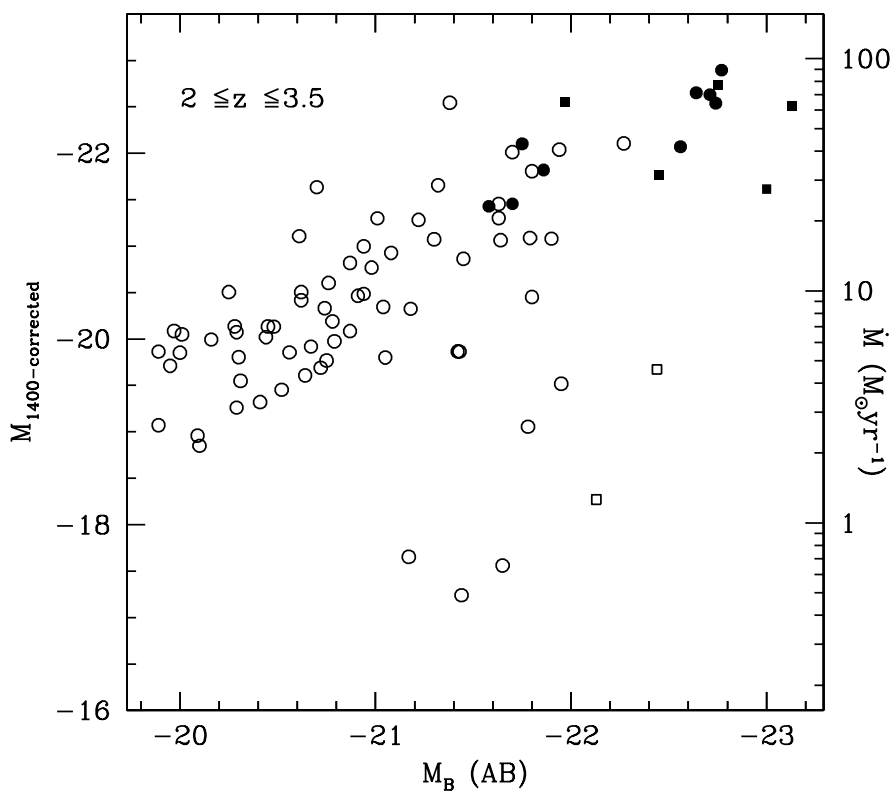}

\caption{ UV de-reddened luminosity at  1400 {\AA}
rest frame vs. observed blue luminosity for  galaxies in the HDF-N (squares)
and  South (circles). A Small Magellanic Cloud extinction curve has been adopted.
Using the Calzetti extinction curve increases the slope of the correlation reaching
$M_{1400}\approx -25$ at $M_B\approx -22.5$.
Filled symbols  represent galaxies  with spectroscopic
redshifts. The corresponding star formation rate (right axis) has been derived
using the Bruzual \& Charlot 2000 code and assuming a Salpeter IMF.
\label{fig2}}
\end{figure}

\end{document}